\documentclass[pre]{revtex4-1}
\usepackage{amsmath,amssymb,bm,xcolor,mathrsfs,graphicx,subfigure,texdraw}
\usepackage{mathtools}
\usepackage{graphicx}
\begin{document}

\title{Fake Superoscillations}

\author{Noemi Barrera}
\affiliation{School of Physics and Astronomy, Raymond and Beverly Sackler Faculty of Exact Sciences, Tel Aviv University}
\author{Eyal Samoi}
\affiliation{Department of Electrical Engineering and Electronics, Holon Institute of Technology (HIT), 52 Golomb St., Holon 5810201, Israel}
\author{Moshe Schwartz}
\affiliation{School of Physics and Astronomy, Raymond and Beverly Sackler Faculty of Exact Sciences, Tel Aviv University}

\begin{abstract}
We construct a signal from "almost" pure oscillations within some low frequency band. We construct it to produce a superoscillation with frequency above the nominal band limit. We find that indeed the required  high frequency is produced but the signal is not a superoscillation as the high frequency components are real and result from an intriguing interference in the frequency domain of the tails of Gaussians concentrated around points within the nominal frequency band.\end{abstract}
\maketitle

\begin{figure}
	\centering
		\includegraphics[width=0.7\textwidth]{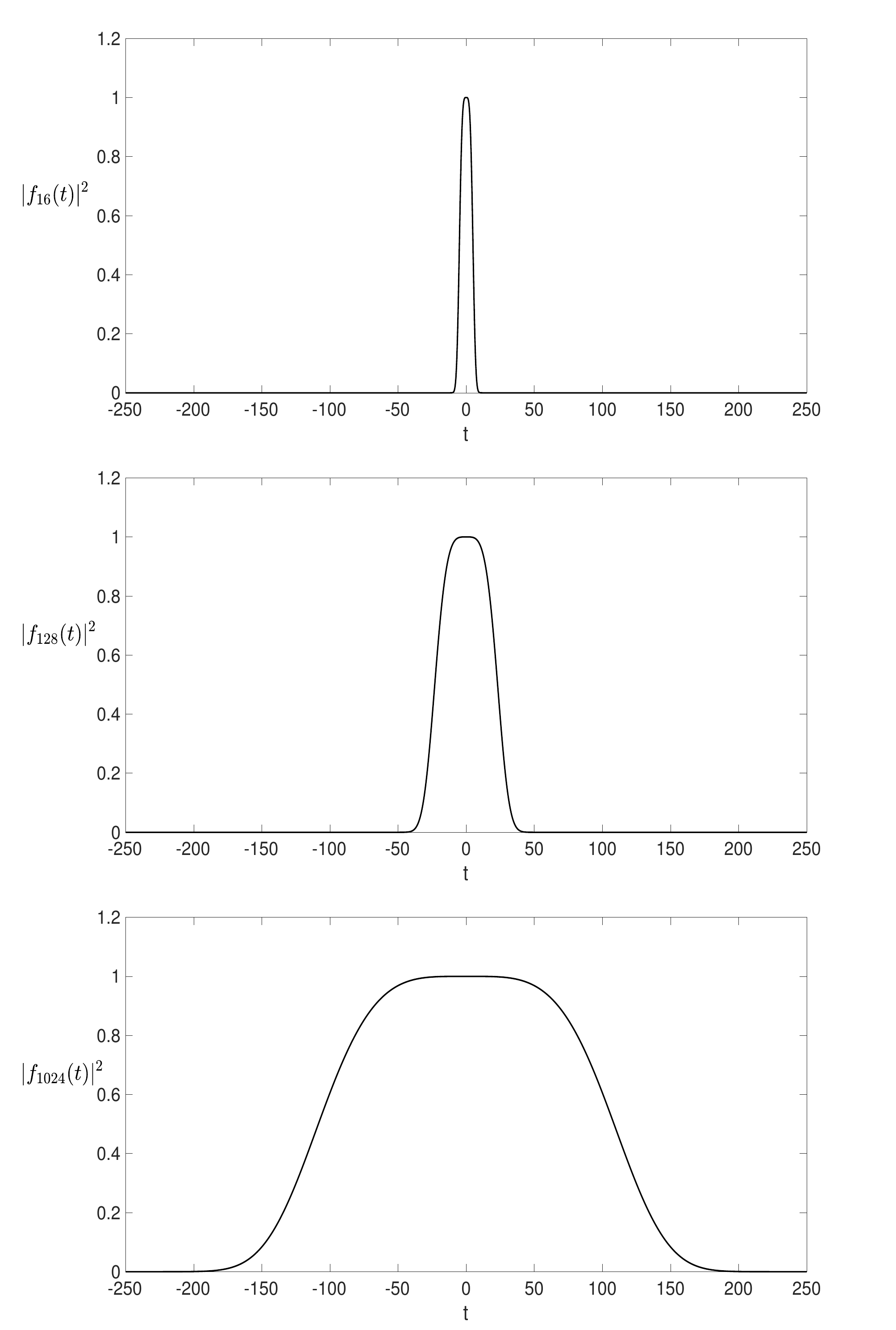}
\caption{The modulus squared of the $f$ signal with $\omega_1/\omega_0=2$ and $n=16,\,128,\,1024$.}\label{fig:modf}
\end{figure}

\begin{figure}
	\centering
		\includegraphics[width=0.7\textwidth]{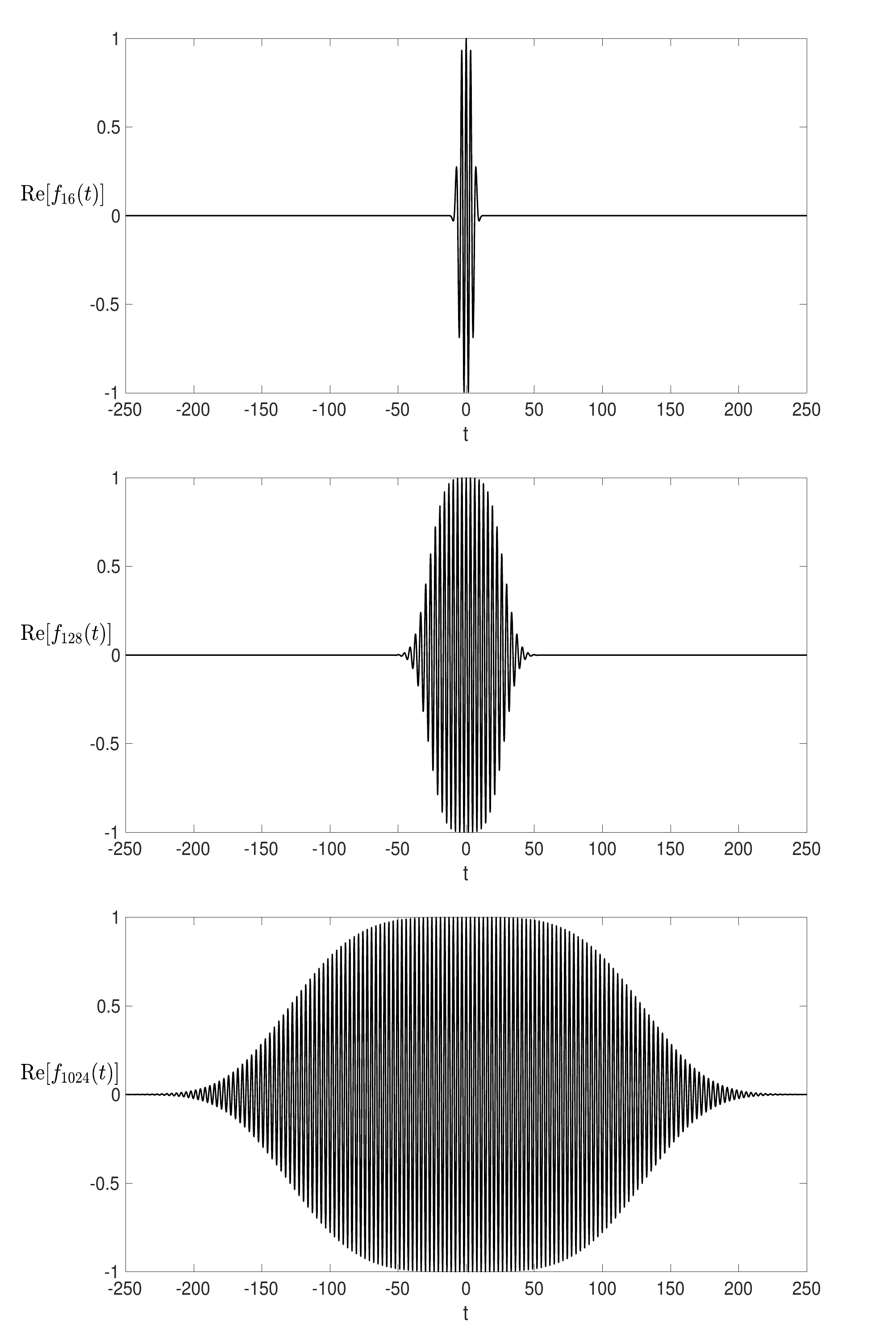}
\caption{The real part of the $f$ signal with $\omega_1/\omega_0=2$ and $n=16,\,128,\,1024$.}\label{fig:f} 
\end{figure}

The phenomenon of superoscillation, introduced into Physics by Aharonov, Popescu and Rohrlich (APR) \cite{APR}, where a band limited signal oscillates locally at frequencies higher than the band limit, seems to offer many beautiful and rather intriguing applications in various fields of science and technology. The list includes, optics \cite{Berry, Zhel, Huang, Lind, Rog, Yuan, Green, Eli, Remez, EliHar, EliBah, Sing}, quantum mechanics \cite{Berry, Sing, AAV, Berry2, Berry3, Kemp, Kemp2, Rein}, signal processing \cite{Ferre, Lee, Lee2} and radar \cite{Wong, Wong2, Wong3}. Yet as is too well known  most of the applications remain at present at the level of science fiction. The main reason is that the wide spread belief that a band limited function cannot oscillate faster than its highest frequency component, is not entirely false. While superoscillation is a proven possibility, it comes at a very high price. The superoscillation yield \cite{Katz}, which is the ratio between the energy going into the fast oscillation and the total energy of the signal, is extremely small. The yield   depends on the ratio between the superoscillation frequency and the band limit frequency and even more strongly on the number of actual oscillations in the superoscillating interval.  The yield is usually too small for most practical purposes.
 Band limited functions are an idealization. In reality any signal must have in its spectrum arbitrarily high frequencies. Thus instead of band limited signals we can talk only about signals poor in frequencies above the band limit .The question that motivated the present article is, can we considerably improve the superoscillation yield by relaxing a bit the requirement that the signal is band limited. In the present article we present results of an extreme attempt to do the impossible of constructing superoscillations without having to pay the price of very small yield. This will lead to the concept of fake superoscillations.  
  Because it will serve later as the basis for the presentation of fake superoscillations, we start our discussion by repeating a number of well-known properties of the superoscillating signal described long ago by  Aharonov Popescu  and Rohrlich (APR) \cite{APR},
\begin{equation}
g_n(t)=\left\{\frac{1}{2}\left[\left(1+\frac{\omega_1}{\omega_0}\exp\left(i\frac{\omega_0}{n}t\right)\right)+\left(1-\frac{\omega_1}{\omega_0}\exp\left(-i\frac{\omega_0}{n}t\right)\right)\right]\right\}^n.
\end{equation}
The almost miraculous behavior of the funtion above stems from two facts: first, it is easily proven by the binomial expansion that $f_n(t)$ is band limited between $-\omega_0$ and $\omega_0$ and, second, that
\begin{equation}
\lim_{n\rightarrow\infty}g_n(t)=\exp(i\omega_1t),
\end{equation}
where $\omega_1/\omega_0$ can be taken as large as we wish. This implies that for given $\omega_0t$ we can find $n$ large enough so that $g_n(t)$ will approximate $\exp(i\omega_1t)$ to a required accuracy. To get a better grip on how the minimal required $n$ depends on $\omega_0t$ and what happens if we fix $n$ and increase $\omega_0t$, it is best to consider first the necessary condition that the square root of the modulus of $g_n(t)$ is close to one. The square of the modulus is given by
\begin{equation}\label{eqn:squareMod}
|g_n(t)|^2=\{1+c\sin^2(\omega_0t/n)\}^n=\{c+1-c\cos^2(\omega_0t/n)\}^n,
\end{equation} 
where $c=(\omega_1/\omega_0)^2-1$.\\
This implies that, for $\omega_1/\omega_0>1$, $1\leq|g_n(t)|^2\leq(\omega_1/\omega_0)^{2n}$.\\
The higher bound is actually attained for $\omega_0t/n=(2m+1)\pi/2$, where $m$ is an integer. This proves that for values of $\omega_0t$ which are of the order of $n$ the modulus suqared is exponentially large in $n$. This implies that, at such values of $\omega_0t$, $g_n(t)$ is very far from approximating $\exp(i\omega_1t)$. Then, when can we expect $g_n(t)$ to approximate $\exp(i\omega_1t)$ reasonably well? Clearly, it must be when $\omega_0t/n\ll1$. In that case, the right hand side of Eq. \ref{eqn:squareMod} can be expanded, so that $|g_n(t)|^2=1+c(\omega_0t)^2/n$ and, by demanding that $|g_n(t)|^2-1<\delta$, we obtain
\begin{equation}\label{eqn:range}
\omega_0t\leq\sqrt{\delta n/c}\equiv\omega_0t^g_{\delta n},
\end{equation}
which gives the time range where $|g_n(t)|^2$ approximates 1 to the required accuracy, $\delta$. So far we have considered a necessary condition. To complete the discussion, consider the local frequency,
\begin{equation}\label{eqn:Arg}
\omega_l(t)=\frac{d}{dt}\textrm{Arg}[g_n(t)]=\frac{d}{dt}\left\{\frac{1}{i}\ln\left[\frac{g_n(t)}{|g_n(t)|}\right]\right\}.
\end{equation}
For the local frequency at $t^g_{\delta n}$ we find
\begin{equation}\label{eqn:om1}
\omega_l(t^g_{\delta n})=\omega_1(1-\delta/n).
\end{equation}
Thus, within the range defined by Eq. \ref{eqn:range} not only the absolute value of the modulus of $g_n(t)$ approximates 1 to the required accuracy but also the complex function $g_n(t)$ approximates $\exp{i\omega_1t}$ to that accuracy. \\
This time range can be made arbitrarily large by increasing $n$ but at the price that beyond that region $|g_n(t)|^2$ becomes exponentially large in $n$. Thus, if we think of the signal carrying energy, the ratio between the useful energy (in the sense of generating superoscillation) to the total energy tends to zero when $n$ tends to infinity. This type of behavior si well known with details that depend on the specific band limited signal chosen to produce the fast local oscililation. It is true that the superoscillation yield can be optimized \cite{Katz} and increased by orders of magnitude in comparison to non-optimized superoscillating signals \cite{Katz2} but it is still mostlyfrustratingly small. It comes to mind that probably the only demand that could be slightly relaxed in order to increase the yield is that the signal should be strictly band limited. After all any realistic band limited signal is not really band limited. There are always frequency components higher than the nominal band limit present, although with very small amplitudes \cite{BerrySUB}.  In fact, band limited signals are non-existing, useful idealizations. \\
Could we gain from that fact? Namely can we construct   a signal that will on one hand have   a high superoscillation yield and on the other will be practically band limited?   In the following we study an extreme example of such a signal. \\
Consider the family of signals, 
\begin{equation}
f_n(t,\alpha)=g_n(t)\exp\{-\alpha c(\omega_0t)^2/2n\},\;\textrm{with}\;\alpha\geq0.
\end{equation}
Clearly
\begin{equation}
\lim_{n\rightarrow\infty}f_n(t,\alpha)=\exp(i\omega_1t).
\end{equation}
Furthermore, it is also clear that for fixed $n$
\begin{equation}
\lim_{n\rightarrow \infty}f_n(t,\alpha)=0\:\textrm{for}\;\alpha>0
\end{equation}
in contrast to the violent large $t$ behavior of $g_n(t)$. From this point on, we will consider only the interesting case $\alpha=1$ and denote $f_n(t,1)$ by $f_n(t)$.\\
Consider $|f_n(t)|^2$. It is easy to show that 
\begin{equation}
\frac{d}{dt}|f_n(t)|^2\leq 0\;\textrm{for}\; t\geq0,
\end{equation}
where the inequality is obeyed as an equality for $t=0$. The short time behavior of the absolute value squared of the signal is given by
\begin{equation}
|f_n(t)|^2=1-\left(\frac{c^2}{2}+\frac{c}{3}\right)\frac{(\omega_0t)^4}{n^3}.
\end{equation}
In accordance with our previous definitio, we define a time range where the departure of $|f_n(t)|^2$ from 1 is less than a prescribed small $\delta$. We find now
\begin{equation}\label{eqn:rangef}
\omega_0t^f_{\delta n}=\left\{\frac{\delta n^3}{c^2/2+c/3}\right\}^{1/4}.
\end{equation}
We define next a $\delta$-dependent yield,
\begin{equation}
Y_\delta=\int_{-t^f_{\delta n}}^{t^f_{\delta n}}dt\,|f_n(t)|^2\Big/\int_{-\infty}^{+\infty}dt\,|f_n(t)|^2,
\end{equation}
which gives the ratio of the energy in the range where $|f_n(t)|^2$ approximates 1 to the required accuracy to the total energy stored in the signal. We could prove now that $Y_\delta$, although dependent, of course, on our tolerance $\delta$, is of order 1 for $\delta$ that could be quite small. Instead, in Figure \ref{fig:modf} we will just show $|f_n(t)|^2$,making this point very clear. The plots presented are for $\omega_1/\omega_0=2$ and for $n=16,\,128,\,1024$. Time is measured in units of $1/\omega_0$.
There results seem strange, because we obtain an extremely high yield, say for $\delta=0.1$. Moreover, the yield seems to improve with $n$. Therefore, we check the local frequency in the range defined by Eq. \ref{eqn:rangef}. It is clear that the local frequency for given $t$ of the signal $f_n(t)$ is identical to that of $g_n(t)$. It has just to be calculated, however, at a different time for the same $\delta$. By following Eq. \ref{eqn:Arg} and \ref{eqn:om1} and by using Eq. \ref{eqn:rangef} we arrive at
\begin{equation}
\omega_l(t^f_{\delta n})=\omega_1\left\{1-\left[\frac{\delta}{(c/2+1/3)n}\right]^{1/2}\right\}.
\end{equation}
It is thus clear  that within the range of the $t^f_{\delta n}$ that correspond more or less to the flat portions of the previous figures ($\delta$ of the order of 0.1 at most) the change in local frequency is very small. This implies that, within those ranges, $f_n(t)$ still approximates $\exp(i\omega_1t)$ to a good accuracy. In Figure \ref{fig:f} we give the real part of the complex signal, $\textrm{Re}[f_n(t)]$ for $n=16,\,128,\,1024$, that clearly show the oscillations. 

\begin{figure}
	\centering
		\includegraphics[width=0.7\textwidth]{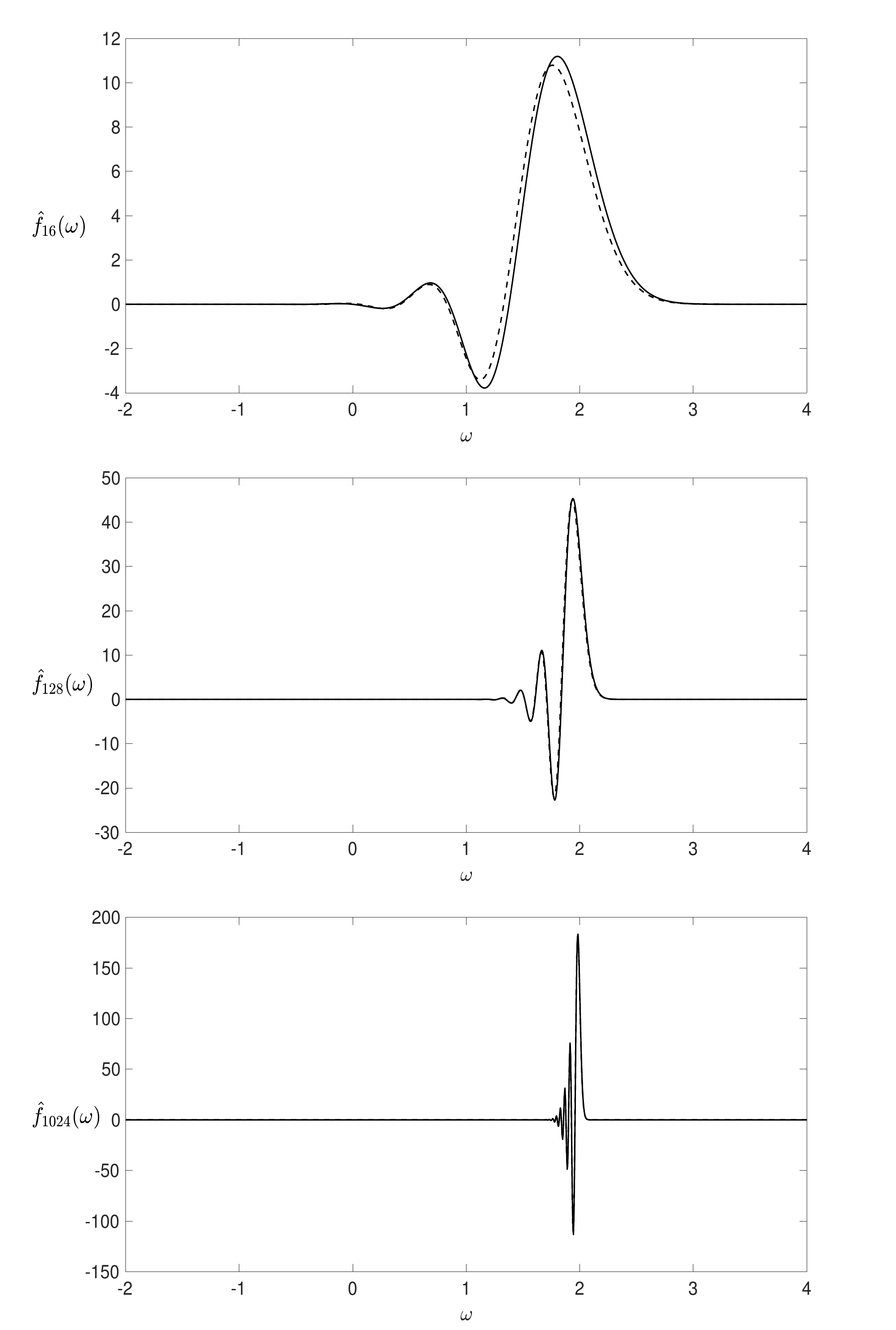}
\caption{Numeric (continuous) and analytic (dashed) approximation of the Fourier transform of $f$ with $\omega_1/\omega_0=2$ and $n=16,\, 128,\, 1024$.} \label{fig:Fourier}
\end{figure}

What we have obtained so far looks vey nice but does not resemble familiar superoscillations at all. To get some more insight consider the Fourier transform of $f_n(t)$,
\begin{equation}\label{eqn:Fourier}
\begin{split}
\hat{f}_n(\omega)=\frac{1}{2^n}\sum_{n=0}^m {{n}\choose{m}}\left(1+i\frac{\omega_1}{\omega_0}\right)^m\times\\
\qquad\qquad\times\left(1-i\frac{\omega_1}{\omega_0}\right)^{n-m}G_n\left(\left(\frac{2m}{n-1}\omega_0-\omega\right)\right)
\end{split}
\end{equation}
where
\begin{eqnarray}
G_n(\eta)&=\frac{1}{2\pi}\int_{-\infty}^{+\infty}dt\,\exp\left[-c\frac{(\omega_0t)^2}{2n}\right]\exp(i\eta t)\nonumber\\
 & =\sqrt\frac{n}{2\pi c\omega_0^2}\exp\left(-\frac{\eta}{2c\omega_0^2}\eta^2\right)
\end{eqnarray}
is a normalized Gaussian. The Fourier transform, $\hat{f}_n(\omega)$, which can be easily shown to be real, is thus a linear combination of Gaussians concentrated around the frequencies $\omega=(2m/(n-1))\omega_0$ with $m$ integers attaining values between 0 and $n$. The width of the Gaussians tends to zero as $n$ tends to infinity. This suggests, at first sight, that the Fourier transforms decay for $\omega$ going out of the range $(-\omega_0,\omega_0)$ and that the decay becomes stronger with $n$. In Figure \ref{fig:Fourier} we present three plots of the Fourier transforms, with $\omega$ in units of $\omega_0$, which contradict our expectations. The continuous line is obtained by numerical integration and the dashed line in an analytic approximation obtained by M. V. Berry \cite{BerryPRIV}
\begin{equation}
\hat{f}_n(\omega)=2\pi\left(\frac{n^2}{c\omega_1/\omega_0}\right)^{1/3}A_i\left[\left(\frac{n^2}{c\omega_1/\omega_0}\frac{\omega-\omega_1}{\omega_0}\right)^{1/3}\right]e^{-\left(\frac{\omega}{\omega_1}-1\right)^2\frac{c+2/3}{4c}n},
\end{equation}
where $A_i$ is the Airy function.
We see that unexpectedly, none of the figures could be considered to be a superoscillation even in the soft "realistic" sense because most of the weight is not concentrated at all between $-\omega_0$ and $\omega_0$ as expected. In fact, $\hat{f}_n(\omega)$ is concentrated mostly near $\omega_1$, with the concentration becoming more pronounced when $n$ is increased. Moreover the structure of $\hat{f}_n(\omega)$ becomes more oscillatory for $\omega$ slightly below $\omega_1$ as $n$ is increased.  The situation is that the high frequencies already exist in the signal. This could have been already guessed from the form of the real part of the signal although not expected from the fact that $\hat{f}_n(\omega)$ is a linear combination of strongly decaying Gaussians concentrated around points in the range $(-\omega_0,\omega_0)$. What we have here is a fascinating interference in the frequency domain. The interference in the range $(-\omega_0,\omega_0)$ is mostly destructive and constructive near $\omega_1$ . The constructive interference  must come from the superposition of the tails of the Gaussians. This may seem, strange, because each of the tails is exponentially small in $n$ at $\omega_1$. We have to recall, however, that the absolute values of pre-factors in front of the Gaussians in equation \ref{eqn:Fourier} are exponentially large in $n$, so the situation is not that surprising. Thus, $f_n(t)$  although constructed along the lines of the APR signal is not a superoscillation at all and we call it a fake superoscillation.\\
To obtain soft superoscillations which are poor in high frequencies yet may look rich in those components locally, we will have to study $f_n(t,\alpha)$ with $\alpha<1$. Recall that $f_n(t,0)=g_n(t)$, which is a hard (strictly band limited) superoscillation and the case studied here with $\alpha=1$  is a fake superoscillation. So, going between those too extremes may prove interesting. By controlling $\alpha$ we will be able to obtain tradeoffs among softness, yield and range of validity of approximating $\exp(i\omega_1t)$.This is postponed, however, to future work. 
To summarize, we introduce a family of signals which are a superposition of "almost" pure oscillations below a certain band limit. The signal is constructed to have a local frequency, $\omega_1$ , larger than the band limit $\omega_0$. The square of the modulus of the signal, its real part and its Fourier transform are obtained. We obtain an interesting interference in the frequency domain that generates real, not just local, oscillation at a frequency close to $\omega_1$. This is due to constructive frequency domain interference of minute tails of Gaussians concentrated around points far away from $\omega_1$ in the range $(-\omega_0,\omega_0)$. These are not superoscillations and we dub such signals fake superoscillations as their construction could have led to the notion that they are really soft super oscillations. The more general family of signals will be hopefully studied in the near future.  \\

\end{document}